\documentclass[longbibliography,physrev,aps,twocolumn]{revtex4-2}
\usepackage{graphicx,epsfig}
\usepackage{times}
\usepackage{amsmath,amssymb,wasysym}
\usepackage{dsfont} % \mathds
\usepackage{color}
\usepackage[linktocpage=true,colorlinks=true, pdfborder={0 0 0},linkcolor=blue,citecolor=blue,urlcolor=blue,anchorcolor=black,linkcolor=blue]{hyperref}

\newcommand{\op}[1]{\hat{#1}}

\newcommand{\normV}[1]{||{#1}||_2}
\newcommand{\normM}[1]{||{#1}||}
\newcommand{\normspec}[1]{||{#1}||_{\text{2}}}

\newcommand{\ident}{\mathds{1}}

\newcommand{\rem}[1]{}

\newcommand{\bra}[1]{\langle#1|}
\newcommand{\ket}[1]{|#1\rangle}

\newcommand{\braket}[2]{\langle#1|#2\rangle}

\newcommand{\sandwich}[3]{\langle#1|#2|#3\rangle}

\newcommand{\normFro}[1]{||#1||_{\text{F}}}
\newcommand{\rca}{\xi}
\newcommand{\rcaa}{\xi_{\text{a}}}
\newcommand{\rcab}{\xi_{\text{b}}}

\DeclareMathOperator{\trace}{Tr}

\newcommand{\Hperturbed}{\op{\cal H}}
\newcommand{\Hs}{\op{H}}
\newcommand{\Hp}{\op{H}_1}
\newcommand{\Hpa}{\op{H}_{1,\text{a}}}
\newcommand{\Hpb}{\op{H}_{1,\text{b}}}
\newcommand{\Ha}{\op{H}_{\text{a}}}
\newcommand{\Hb}{\op{H}_{\text{b}}}
\newcommand{\oK}{\op{K}}
\newcommand{\oKp}{\oK_1}
\newcommand{\Na}{\op{N}_{\text{a}}}
\newcommand{\Nb}{\op{N}_{\text{b}}}
\newcommand{\ga}{g_{\text{a}}}
\newcommand{\gb}{g_{\text{b}}}
\newcommand{\alphaa}{\alpha_{\text{a}}}
\newcommand{\alphab}{\alpha_{\text{b}}}

\newcommand{\order}{n}
\newcommand{\Order}{n}
\newcommand{\Ordera}{\Order_{\text{a}}}
\newcommand{\Orderb}{\Order_{\text{b}}}

\newcommand{\PT}{${\cal{PT}}$}
\newcommand{\ev}{E}
\newcommand{\evEP}{\ev_{\text{EP}}}

\newcommand{\state}{\psi}
\newcommand{\stateEP}{\state_{\text{EP}}}
\newcommand{\stateEPa}{\state_{\text{EPa}}}
\newcommand{\stateEPb}{\state_{\text{EPb}}}
\newcommand{\Jordan}[1]{j_{#1}}
\newcommand{\Jordana}[1]{j_{\text{a},#1}}
\newcommand{\Jordanb}[1]{j_{\text{b},#1}}
\newcommand{\GF}{\op{G}}

\newcommand{\epsmp}{\varepsilon_{\text{mp}}}

\newcommand{\HL}[1]{#1}

\begin{document}

\title{Revisiting the hierarchical construction of higher-order exceptional points}
\author{Jan Wiersig}
\affiliation{Institut f{\"u}r Physik, Otto-von-Guericke-Universit{\"a}t Magdeburg, Postfach 4120, D-39016 Magdeburg, Germany}
\email{jan.wiersig@ovgu.de}
\date{\today}
\begin{abstract}
Higher-order exceptional points in the spectrum of non-Hermitian Hamiltonians describing open quantum or wave systems have a variety of potential applications in particular in optics and photonics. However, the experimental realization is notoriously difficult. Recently, Q. Zhong {\it et al.} [Phys. Rev. Lett. 125, 203602 (2020)] have introduced a robust construction where a unidirectional coupling of two subsystems having exceptional points of the same order leads generically to a single exceptional point of twice the order. 
Here, we investigate this scheme in a different manner by exploiting the nilpotency of the traceless part of the involved Hamiltonians. We generalize the scheme and derive a simple formula for the spectral response strength of the composite system hosting a higher-order exceptional point. Its relation to the spectral response strengths of the subsystems is discussed. Moreover, we investigate nongeneric perturbations. The results are illustrated with an example.
\end{abstract}
%\pacs{42.25.-p, 42.60.Da, 42.25.Dd}
% 42.25.-p Wave optics, includes:
%          42.25.Gy Edge and boundary effects; reflection and refraction
%          42.25.Fx Diffraction and scattering
%          42.25.Hz Interference
% 42.55.Sa Microcavity and microdisk lasers
% 42.60.Da Resonators, cavities, amplifiers, arrays, and rings
% 42.25.Dd Wave propagation in random media
% 05.45.Mt Quantum chaos; semiclassical methods
\maketitle

\section{Introduction}
\label{sec:intro}
The recent years have witnessed an enormous progress in the understanding and the experimental realization of open quantum systems and open classical wave systems. This has stimulated a paradigmatic shift in that dissipation should not always be considered as a bug but as a feature. These potentially useful, physically interesting, and often counterintuitive effects of dissipation are particularly prominent near special degeneracies called exceptional points (EPs)~\cite{MA19,ORN19}. At an EP of order~$\order$ (EP$_\order$) exactly $\order$ eigenvalues and the corresponding eigenstates of the effective Hamiltonian $\op{H}$ coalesce~\cite{Kato66,Heiss00,Berry04,Heiss04}. This is in strong contrast to conventional degeneracies, so-called diabolic points~\cite{BW84}, where only the eigenvalues coalesce. The existence of an EP requires $\op{H}$ to be non-normal, $[\op{H},\op{H}^\dagger] \neq 0$. This is a stronger condition than non-Hermiticity, $\op{H} \neq \op{H}^\dagger$. 
% existence 
EPs are not just curious mathematical objects, they have been realized experimentally, first in microwave cavities~\cite{DGH01,DDG04}, later in optical microcavities~\cite{LYM09,POL14,POL16}, coupled atom-cavity composites~\cite{CKL10}, photonic lattices~\cite{RBM12}, semiconductor exciton-polariton systems~\cite{GEB15}, and ultrasonic cavities~\cite{SKM16}. 

% EP-sensitivity
When a Hamiltonian $\Hs$ with an EP$_\order$ is subjected to a small perturbation of strength~$\varepsilon > 0$,
\begin{equation}\label{eq:H}
\Hperturbed = \Hs+\varepsilon\Hp \ ,
\end{equation}
then the resulting energy (or frequency) splittings are typically proportional to the $\order$th root of $\varepsilon$~\cite{Kato66}. For sufficiently small perturbations this is larger than the linear scaling near a diabolic point, which can be exploited for sensing applications~\cite{Wiersig14b,Wiersig20b,Wiersig20c}. A number of experiments have demonstrated the feasibility of EP-based sensors~\cite{COZ17,HHW17,ZCZ18,CSH18,DLY19,ZSL19,LLS19,HSC19,PNC20,KCE22}.
The enhanced response can be quantified with a single quantity, the spectral response strength~$\rca$~\cite{Wiersig22,Wiersig22b}.

% exceptional surfaces
The high sensitivity of EPs has also a drawback: Fabrication requires a delicate fine tuning of experimental parameters, in particular for higher-order EPs. One solution to this problem are exceptional surfaces~\cite{ZRK19,ZNO19} embedded in a higher-dimensional parameter space. Robust EP-based sensing can be achieved if the system’s response is tailored such that a large class of fabrication errors and experimental uncertainties shift the operation point along the exceptional surface. Perturbations that drive the parameters away from the surfaces cause an EP-enhanced energy splitting. Exceptional surfaces have been suggested for optical amplifiers~\cite{ZOE20}, sensing~\cite{QXZ21}, control of spontaneous emission~\cite{ZHO21}, and chiral perfect absorbers~\cite{SZM22}.

% hierarchical construction of higher-order EPs
Another rather robust method to obtain higher-order EPs is the hierarchical construction~\cite{ZKO20}: two subsystems, each with an EP of order $\order$, are coupled in a unidirectional way resulting in a composite system with an EP of order $2\order$. This method provides a realistic route to higher-order EP-based devices. 
% aim of this paper
The aim of this paper is to revisit the hierarchical construction of higher-order EPs. Using a different mathematical approach allows us to generalize the scheme, to determine the spectral response strength, and to relate the latter to the response strengths of the two subsystems. 

% outline of the paper
The outline of the paper is as follows. In Sec.~\ref{sec:hc} the concept of hierarchical construction of higher-order EPs is briefly explained. Section~\ref{sec:np} presents a different point of view and an extension. In Sec.~\ref{sec:rca} the relation to the spectral response strength is discussed. Unidirectional-coupling preserving perturbations are investigated in Sec.~\ref{sec:ucpp}. An illustrative example is considered in Sec.~\ref{sec:examples} and a summary is given in Sec.~\ref{sec:summary}.

\section{Hierarchical construction}
\label{sec:hc}
In this section we shortly review the hierarchical construction of higher-order EPs as developed in Ref.~\cite{ZKO20}. Starting from two $\Order\times\Order$ Hamiltonians $\Ha$ and $\Hb$ each at an EP of order $\Order$ with the \emph{same} eigenvalue $\evEP$ the $2\Order\times 2\Order$ Hamiltonian 
\begin{equation}\label{eq:HB1}
\op{H} = \left(\begin{array}{cc}
\Ha &  0 \\
\oK & \Hb\\
\end{array}\right) 
\end{equation}
is constructed. Assuming a generic choice of the $\Order\times\Order$ coupling matrix~$\oK$ the authors of Ref.~\cite{ZKO20} have shown by carefully investigating the eigenvalue equation of $\op{H}$ that it has an EP of order $2\Order$.
The assumed generic choice of the unidirectional coupling $\oK$ had been interpreted as a disorder-immune behavior of the construction. 

As a concrete experimental setup it was suggested to use two evanescently coupled optical microrings each supporting two modes, one traveling clockwise and one counterclockwise~\cite{ZKO20}. One of the microrings exhibits gain, the other one exhibits an equal amount of loss. The unidirectional coupling can be achieved by evanescently coupling the lossy microring to a semi-infinite waveguide with an end mirror. In this way an EP$_4$ is implemented in a robust manner.

Note that a unidirectional coupling in Eq.~(\ref{eq:HB1}) does not necessarily violate Lorentz reciprocity. It is true that for reciprocal systems there is an orthonormal basis in which the Hamiltonian~$\op{H}$ is represented by a complex-symmetric matrix~\cite{DP68,SPK02}. This basis consists of standing waves, which are invariant under time reversal. In a different basis, however, the same Hamiltonian~$\op{H}$ can be represented by an asymmetric matrix. For instance, fully asymmetric backscattering of counterpropagating waves in whispering-gallery cavities appears in a traveling-wave basis as a unidirectional coupling, even if the system is Lorentz  reciprocal~\cite{Wiersig11,POL16,Wiersig18b}. Other possibilities to introduce a unidirectional coupling are via reservoir engineering~\cite{MC15} and via space-time modulation~\cite{LJC22}.
If one considers Liouvillian EPs~\cite{MMC19,Wiersig20} then a unidirectional coupling in the matrix representation of the Liouville operator can result from quantum jump terms in the master equation of the open quantum system, see, e.g., the supplementary material of Ref.~\cite{CAJ21}.
 
\section{Exploiting the nilpotency}
\label{sec:np}
\HL{An $\order\times \order$ Hamiltonian~$\op{H}$ at an EP$_\order$ has an eigenvalue $\evEP$ with algebraic multiplicity $\order$ but only one eigenvector (the  geometric multiplicity is 1). Hence, the trace of $\op{H}$ is $\trace{\op{H}} = \order\evEP$. We introduce the traceless part of $\op{H}$,
\begin{equation}\label{eq:N}
	\op{N} := \op{H}-\evEP\ident \ ,
\end{equation}
which exhibits the same properties as $\op{H}$ except that all eigenvalues are shifted to zero. Here and henceforth $\ident$ is the identity with proper dimensions in the particular context. Although not obvious here, it is known that $\op{N}$ is nilpotent of index $\order$, i.e., $\op{N}^\order = 0$ but $\op{N}^{\order-1} \neq 0$; see, e.g., Refs.~\cite{Kato66,TE05,Wiersig22,HJ13}. The property $\op{N}^\order = 0$ ensures that all eigenvalues of $\op{N}$ are zero and $\op{N}^{\order-1} \neq 0$ ensures here that only one eigenvector of $\op{N}$ exists.
The above statement is also valid in the other direction: An $\order\times \order$ Hamiltonian $\op{H} = \evEP\ident + \op{N}$ with $\op{N}$ being nilpotent of index~$\order$ has an EP$_\order$ with eigenvalue $\evEP$.}

In the following we exploit the nilpotency to gain insight into the hierarchical construction of higher-order EPs. For didactic reasons we first restrict ourselves to the merging of two EP$_2$. For the traceless part of the Hamiltonians $\Ha$ and $\Hb$ then holds $\Na^2 = 0$ and $\Nb^2 = 0$. Plugging this into Eq.~(\ref{eq:HB1}) we get for the traceless part of $\op{H}$
\begin{equation}\label{eq:NB1}
\op{N} = \left(\begin{array}{cc}
\Na &  0 \\
\oK & \Nb\\
\end{array}\right) \ ,
\end{equation}
\begin{equation}\label{eq:NB2}
\op{N}^2 = \left(\begin{array}{cc}
0 &  0 \\
\oK\Na+\Nb\oK & 0\\
\end{array}\right) \ ,
\end{equation}
\begin{equation}\label{eq:NB3}
\op{N}^3 = \left(\begin{array}{cc}
0 &  0 \\
\Nb\oK\Na & 0\\
\end{array}\right) \ ,
\end{equation}
and $\op{N}^4 = 0$. Hence, $\op{H}$ has an EP$_4$ if and only if $\Nb\oK\Na \neq 0$. In the special case $\Nb\oK\Na = 0$, $\op{H}$ has an EP$_3$ if $\oK\Na+\Nb\oK\neq 0$, otherwise it has an EP$_2$. 

This scheme can be easily extended not only to higher order but also to merging EPs of different order, thereby going beyond Ref.~\cite{ZKO20}. To do so, we consider here the general situation where $\Ha$ has an EP of order $\Ordera\geq 2$ and $\Hb$ has an EP of order $\Orderb\geq 2$. The matrix dimensions are $\Ordera\times\Ordera$ for $\Ha$, $\Orderb\times\Orderb$ for $\Hb$, $\Orderb\times\Ordera$ for $\oK$, and consequently $(\Ordera+\Orderb)\times(\Ordera+\Orderb)$ for $\op{H}$. With $\Na^{\Ordera} = 0$ and $\Nb^{\Orderb} = 0$ it is straightforward to show that 
\begin{equation}\label{eq:NB4}
\op{N}^{\Ordera+\Orderb-1} = \left(\begin{array}{cc}
0 &  0 \\
\Nb^{\Orderb-1}\oK\Na^{\Ordera-1} & 0\\
\end{array}\right) 
\end{equation}
and $\op{N}^{\Ordera+\Orderb} = 0$. Hence the Hamiltonian $\op{H}$ has an EP of order $\Ordera+\Orderb$ provided that
\begin{equation}\label{eq:generic}
\Nb^{\Orderb-1}\oK\Na^{\Ordera-1} \neq 0 \ .
\end{equation}
Clearly, this condition is violated for special coupling matrices, such as $\oK = \Na$ or $\oK = \Nb$, but it holds for the generic situation. However, what if $\oK$ is close to a nongeneric situation? This question was not addressed in Ref.~\cite{ZKO20}. We can study it by using the concept of spectral response strength.

\section{Spectral response strength}
\label{sec:rca}
We consider again a system described by an $\order\times\order$ Hamiltonian~$\op{H}$ hosting an EP$_\order$ with eigenvalue $\evEP$. In Ref.~\cite{Wiersig22} it has been shown that the system's spectral response to perturbations can be characterized by a single quantity, the spectral response strength 
\begin{equation}\label{eq:rca}
\rca = \normspec{\op{N}^{\order-1}} = \normFro{\op{N}^{\order-1}} \ ,
\end{equation}
with the traceless part of the Hamiltonian~$\op{N}$ in Eq.~(\ref{eq:N}). Here $\normspec{\cdot}$ is the spectral norm of a matrix $\op{A}$ (see, e.g., Ref.~\cite{HJ13})
\begin{equation}\label{eq:defspn}
\normspec{\op{A}} := \max_{\normV{\psi} = 1}\normV{\op{A}\psi} \ ,
\end{equation}
with the vector 2-norm $\normV{\psi} = \sqrt{\braket{\psi}{\psi}}$ of a vector $\ket{\psi}$ based on the usual inner product in complex vector space.
$\normFro{\cdot}$ is the Frobenius norm
\begin{equation}\label{eq:Fronorm}
\normFro{\op{A}} := \sqrt{\trace{(\op{A}^\dagger\op{A})}}  = \sqrt{\sum_{ij}|A_{ij}|^2} 
\end{equation}
where $A_{ij}$ are the matrix elements of $\op{A}$ in any orthonormal basis. \HL{The following inequalities hold~\cite{Johnston21}
\begin{equation}\label{eq:2F2}
\normspec{\op{A}} \leq 	\normFro{\op{A}} \leq \sqrt{r} \normspec{\op{A}} \ ,
\end{equation}	
where $r$ is the rank of $\op{A}$. Hence, in the special case of a rank-1 matrix both matrix norms give the same result. 
This happens in Eq.~(\ref{eq:rca}) where $\op{N}^{\order-1}$ has rank one as the matrix $\op{N}$ is nilpotent of index $\order$, see Ref.~\cite{Wiersig22}}. The spectral response strength~$\rca$ [Eq.~(\ref{eq:rca})] shows up as a factor in the bound of the change of the eigenvalues~$\ev_j$ of the perturbed Hamiltonian in Eq.~(\ref{eq:H})
\begin{equation}\label{eq:specresponse}
|\ev_j-\evEP|^\order \leq \varepsilon \normspec{\Hp}\,\rca  \ .
\end{equation}
Note that $\normspec{\Hp}$ and $\normFro{\Hp}$ are in general not equal but related by the inequalities~(\ref{eq:2F2}).
A large $\rca$ indicates a strong spectral response to generic perturbations in terms of large energy splittings. It can therefore be used to distinguish the spectral response of two EPs of the same order. Immediate applications of this quantity are (i) assessing the suitability of a design of an optical structure for EP-based applications (calculating $\rca$ in the discussed context is easy and can often been done analytically even for higher-order EPs) and (ii) finding strong-spectral-response regions within an exceptional surface.

The spectral response strength of the hierarchically constructed Hamiltonian in Eq.~(\ref{eq:HB1}) is according to Eq.~(\ref{eq:rca}) $\rca = \normspec{\op{N}^{\Ordera+\Orderb-1}}$. Note that because of the fact that $\op{N}^{\Ordera+\Orderb-1}$ has rank 1 one can here use also the Frobenius norm to compute $\rca$. With Eq.~(\ref{eq:NB4}) one can easily derive the important result
\begin{equation}\label{eq:rcafull}
\rca = \normspec{\Nb^{\Orderb-1}\oK\Na^{\Ordera-1}} = \normFro{\Nb^{\Orderb-1}\oK\Na^{\Ordera-1}} \ .
\end{equation}
In the generic situation [see inequality~(\ref{eq:generic})] this quantity is nonzero, but it can be small close to a nongeneric situation. In this sense Eq.~(\ref{eq:rcafull}) provides a straightforward way to reveal the closeness to a nongeneric situation.

\subsection{Upper bound}
We can derive an upper bound for the response strength of the composite system in Eq.~(\ref{eq:rcafull}) by using the submultiplicativity of the spectral norm, i.e.,
\begin{equation}\label{eq:submul}
\normM{\op{A}\op{B}} \leq \normM{\op{A}}\,\normM{\op{B}} 
\end{equation}
for all matrices $\op{A}$ and $\op{B}$, see, e.g., Ref.~\cite{HJ13}.
With the spectral response strength
\begin{equation}\label{eq:rcaa}
\rcaa = \normspec{\Na^{\Ordera-1}}
\end{equation}
associated with the Hamiltonian $\Ha$ and the spectral response strength
\begin{equation}\label{eq:rcab}
\rcab = \normspec{\Nb^{\Orderb-1}}
\end{equation}
associated with the Hamiltonian $\Hb$ it follows directly from the submultiplicativity in inequality~(\ref{eq:submul})
\begin{equation}\label{eq:rcabound}
\rca \leq \rcaa\rcab\normspec{\oK} \ .
\end{equation}
Inequality~(\ref{eq:rcabound}) implies that a necessary condition for a strong response strength of the hierarchically constructed higher-order EP is that the upper bound $\rcaa\rcab\normspec{\oK}$ is large. Hence, the original lower-order EPs should each have a large response strength and the coupling of the two EPs should be strong enough in terms of $\normspec{\oK}$.
Note that if the matrix $\oK$ contains only one nonzero matrix element then this matrix is of rank one. For such matrices it follows again from the inequalities~(\ref{eq:2F2}) that  $\normspec{\oK} = \normFro{\oK}$. Hence, we can use here the Frobenius norm which is easier to calculate. The individual response strengths $\rcaa$ and $\rcab$ can always be determined by using the Frobenius norm.

\subsection{Coupling amplitude}
\label{sec:cme}
We consider the linearly independent Jordan vectors $\ket{\Jordan{1}}, \ldots, \ket{\Jordan{\order}}$ of an EP$_\order$ defined by the Jordan chain (see, e.g., Ref.~\cite{SM03})
\begin{eqnarray}\label{eq:jc1}
\op{N}\ket{\Jordan{1}} & = & 0 \ ,\\ 
\label{eq:jc}
\op{N}\ket{\Jordan{l}} & = & \ket{\Jordan{l-1}} 
\;;\, l = 2,\ldots,\order 
\end{eqnarray}
with the operator $\op{N}$ from Eq.~(\ref{eq:N}). Only $\ket{\Jordan{1}}$ is an eigenstate of the Hamiltonian, which is $\ket{\Jordan{1}} = \ket{\stateEP}$. The Jordan vectors are not uniquely determined by Eqs.~(\ref{eq:jc1}) and (\ref{eq:jc}). This can be fixed by the following conditions (see, e.g., Ref.~\cite{Wiersig22})
\begin{eqnarray}\label{eq:ortho1}
\braket{\Jordan{1}}{\Jordan{1}} & = & 1 \ ,\\
\label{eq:ortho2}
\braket{\Jordan{\order}}{\Jordan{l}} & = & 0
\quad\mbox{for}\; l = 1, \ldots, \order-1 \ .
\end{eqnarray}
With this normalization and orthogonalization it was shown in Ref.~\cite{Wiersig22} that the spectral response strength can be expressed by the length of the ``last Jordan vector'',
\begin{equation}\label{eq:rcajordan}
\rca = \frac{1}{\normspec{\Jordan{\order}}} \ .
\end{equation}
For our two different EPs we introduce the normalized EP eigenstates $\ket{\stateEPa}$, $\ket{\stateEPb}$ and Jordan vectors $\{\ket{\Jordana{1}},\ldots,\ket{\Jordana{\Ordera}}\}$ and $\{\ket{\Jordanb{1}},\ldots,\ket{\Jordanb{\Orderb}}\}$ in the corresponding subspaces. There we can write according to Eqs.~(\ref{eq:jc1})-(\ref{eq:ortho2}) 
\begin{equation}
\Na^{\Ordera-1} = \frac{\ket{\stateEPa}\bra{\Jordana{\Ordera}}}{\normspec{\Jordana{\Ordera}}^2}
\end{equation}
and
\begin{equation}
\Nb^{\Orderb-1} = \frac{\ket{\stateEPb}\bra{\Jordanb{\Orderb}}}{\normspec{\Jordanb{\Orderb}}^2} \ .
\end{equation}
This gives
\begin{equation}
\Nb^{\Orderb-1}\oK\Na^{\Ordera-1} = \frac{\ket{\stateEPb}\bra{\Jordanb{\Orderb}}\oK\ket{\stateEPa}\bra{\Jordana{\Ordera}}}{\normspec{\Jordanb{\Orderb}}^2\normspec{\Jordana{\Ordera}}^2} \ .
\end{equation}
Using the Frobenius norm [Eq.~(\ref{eq:Fronorm})] in Eq.~(\ref{eq:rcafull}) we obtain
\begin{equation}\label{eq:rcatrace}
\rca = \frac{\sqrt{\trace{\left(\ket{\Jordana{\Ordera}}|\bra{\Jordanb{\Orderb}}\oK\ket{\stateEPa}|^2\bra{\Jordana{\Ordera}}\right)}}}{\normspec{\Jordanb{\Orderb}}^2\normspec{\Jordana{\Ordera}}^2} \ .
\end{equation}
We define the two unit vectors
\begin{eqnarray}\label{eq:normja}
\ket{\widetilde{\Jordana{\Ordera}}} & := & \frac{\ket{\Jordana{\Ordera}}}{\normV{\Jordana{\Ordera}}} \ ,\\
\label{eq:normjb}
\ket{\widetilde{\Jordanb{\Orderb}}} & := & \frac{\ket{\Jordanb{\Orderb}}}{\normV{\Jordanb{\Orderb}}} \ .
\end{eqnarray}
Evaluating the trace in Eq.~(\ref{eq:rcatrace}) in an orthonormal basis in the subspace corresponding to the first EP ($\Ha$) with one element being the unit vector in Eq.~(\ref{eq:normja}) gives the result
\begin{equation}\label{eq:rcacouplingK}
\rca = \rcaa\rcab|\bra{\widetilde{\Jordanb{\Orderb}}}\oK\ket{\stateEPa}| \ ,
\end{equation}
where we have also utilized Eqs.~(\ref{eq:rcaa}), (\ref{eq:rcab}), (\ref{eq:rcajordan}), and~(\ref{eq:normjb}). The more specific Eq.~(\ref{eq:rcacouplingK}) is fully consistent with the inequality~(\ref{eq:rcabound}). 
It allows us to determine the response strength of the resulting EP in terms of the response strengths of the individual EPs, $\rcaa$ and $\rcab$, and the coupling amplitude $\bra{\widetilde{\Jordanb{\Orderb}}}\oK\ket{\stateEPa}$. \HL{Interestingly, the latter is not symmetric with respect to the two subsystems a and b. This asymmetry originates from the unidirectional coupling of the two subsystems.

Another benefit of Eq.~(\ref{eq:rcacouplingK}) is that it provides a possibility to design the unidirectional coupling in order to maximize the spectral response at the resulting higher-order EP. For fixed properties of the original lower-order EPs and fixed total coupling strength, measured e.g. by $\normspec{\oK}$, the coupling matrix $\oK$ can be chosen such that $|\bra{\widetilde{\Jordanb{\Orderb}}}\oK\ket{\stateEPa}|$ is maximized.
}

\section{Unidirectional-coupling preserving perturbations}
\label{sec:ucpp}
In this section we show that all unidirectional-coupling preserving perturbations
\begin{equation}\label{eq:HBp}
\Hp = \left(\begin{array}{cc}
\Hpa &  0 \\
\oKp & \Hpb\\
\end{array}\right) 
\end{equation}
are nongeneric in the sense that they do not lead to the naively expected $\order$th-root splitting of the eigenvalues of the Hamiltonian at an EP$_\order$ with $\order = \Ordera+\Orderb$. 
The highest-order generic contribution to the energy splitting is
\begin{equation}\label{eq:splitting}
(\ev_j-\evEP)^\order = \varepsilon\trace{\left(\op{N}^{\order-1}\Hp\right)} \ .
\end{equation}
This equation is proven in the appendix.

From Eq.~(\ref{eq:NB4}) with $\order = \Ordera+\Orderb$ we can derive
\begin{equation}\label{eq:HBp1}
\op{N}^{\order-1}\Hp = \left(\begin{array}{cc}
0 &  0 \\
\Nb^{\Orderb-1}\oK\Na^{\Ordera-1}\Hpa & 0\\
\end{array}\right) \ .
\end{equation}
From the obvious fact $\trace{(\op{N}^{\order-1}\Hp)} = 0$ it follows that the highest-order contribution of the energy splitting in Eq.~(\ref{eq:splitting}) is zero. Hence, the rather general looking perturbation in Eq.~(\ref{eq:HBp}), which perturbs the two subsystems and the unidirectional coupling in the most general way, but preserves the unidirectional-coupling structure, cannot take advantage of the full sensitivity of the EP$_\order$.

Note the special case of Eq.~(\ref{eq:HBp}) with $\Hpa = 0 = \Hpb$ is a perturbation that leaves the system on an EP$_\order$ if inequality~(\ref{eq:generic}) holds with $\oK\to\oK+\oKp$. It is an example of an exceptional surface. 

\section{Example}
\label{sec:examples}
\subsection{The system}
% dimer
As an example we consider the unidirectional coupling of a parity-time (\PT) symmetric dimer to a \PT-symmetric trimer. The Hamiltonian of the dimer is
\begin{equation}\label{eq:HsPT}
\Ha = \left(\begin{array}{cc}
\omega_0+i\alphaa & \ga\\
\ga   & \omega_0-i\alphaa\\
\end{array}\right) \ .
\end{equation}
The real-valued quantity $\omega_0$ is the frequency, $\alphaa > 0$ the gain/loss coefficient, and $\ga > 0$ the coupling strength. Possible realizations of this system are composed of two coupled waveguides~\cite{RMG10} or resonators~\cite{HHW17}. It is a \PT-symmetric system as it is invariant under the combined action of parity (exchange of waveguide/resonators) and time-reversal (exchange of gain and loss) operations. If $\ga = \alphaa$ the dimer has an EP of order $\Ordera = 2$ with eigenvalue $\evEP = \omega_0$. The response strength according to Eqs.~(\ref{eq:N}) and~(\ref{eq:rca}) has been calculated in Ref.~\cite{Wiersig22}, 
\begin{equation}\label{eq:examplercaa}
\rcaa = 2\ga \ .
\end{equation}

% trimer
The Hamiltonian of the trimer is 
\begin{equation}\label{eq:H0EPd}
\Hb = \left(\begin{array}{ccc}
\omega_0 + i\alphab & \gb & 0 \\
\gb & \omega_0 & \gb\\
0 & \gb & \omega_0 -i\alphab\\
\end{array}\right) \ .
\end{equation}
Again, $\omega_0$ is the real-valued frequency, $\alphab > 0$ is the gain/loss coefficient, and $\gb > 0$ is the coupling strength. If $\alphab = \sqrt{2}\gb$ the trimer is at an EP of order $\Orderb = 3$ with eigenvalue $\evEP = \omega_0$. The response strength is~\cite{Wiersig22}
\begin{equation}\label{eq:examplercab}
\rcab = 4\gb^2 \ .
\end{equation}
Possible realizations in terms of coupled resonators and waveguides are discussed in Refs.~\cite{DG12,HHW17}.

\begin{figure}[ht]
\includegraphics[width=0.7\columnwidth]{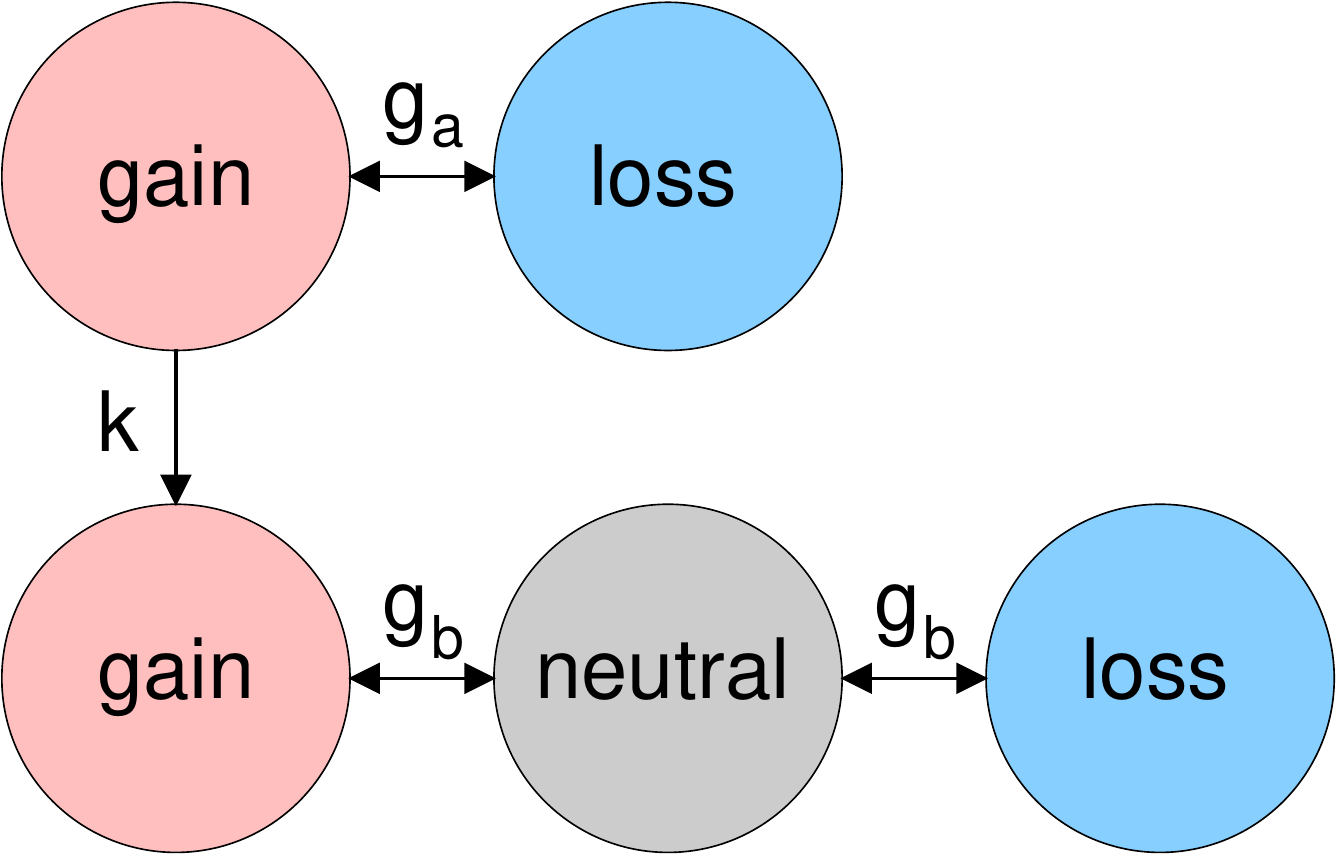}
\caption{Sketch of the unidirectionally-coupled \PT-symmetric dimer and trimer. Each disk represents a single-mode cavity. The coupling strengths $\ga$ and $\gb$ couple the modes in a symmetric manner. $k$ is the strength of the unidirectional coupling of the gainy resonator in the dimer to the gainy resonator in the trimer.}
\label{fig:example}
\end{figure}

% unidirectional coupling
We couple the two subsystems with EPs of different order by a unidirectional coupling from the gainy resonator of the dimer to the gainy resonator of the trimer, see Fig.~\ref{fig:example}. The coupling matrix is
\begin{equation}\label{eq:exampleK}
\oK = \left(\begin{array}{cc}
k & 0 \\
0 & 0 \\
0 & 0 \\
\end{array}\right) \ .
\end{equation}
For illustration purposes, we have employed such a simple coupling matrix. Note that the theory is not restricted to sparse coupling matrices.

Using $\Ha$, $\Hb$, and $\oK$ in the full Hamiltonian in Eq.~(\ref{eq:HB1}) and the nilpotent matrix in Eq.~(\ref{eq:N}) allows us to calculate Eq.~(\ref{eq:NB4}) with
\begin{equation}\label{eq:C}
\Nb^{\Orderb-1}\oK\Na^{\Ordera-1} = k\ga\gb^2\left(\begin{array}{cc}
-i & -1 \\
-\sqrt{2} & i\sqrt{2} \\
i & 1 \\
\end{array}\right)\ .
\end{equation}
This matrix is nonzero for nonvanishing coupling strengths $k$, $\ga$, and $\gb$. Hence, the full system of coupled dimer and trimer has an EP of order $\order = \Ordera+\Orderb = 5$. Using Eqs.~(\ref{eq:rcafull}) and~(\ref{eq:C}) the response strength associated with this EP is 
\begin{equation}\label{eq:examplerca}
\rca = \sqrt{8}|k|\ga\gb^2 \ .
\end{equation}
The bound in the inequality~(\ref{eq:rcabound}) is fulfilled as one can see by first calculating $\normspec{\oK}$, which can also be evaluated in the Frobenius norm as $\oK$ here has rank one. We immediately get $\normspec{\oK} = |k|$. With the individual response strengths in Eqs.~(\ref{eq:examplercaa}) and~(\ref{eq:examplercab}) we obtain $\rca \leq 8|k|\ga\gb^2$ which is fully consistent with Eq.~(\ref{eq:examplerca}). 

% check Eq.~(\ref{eq:rcacouplingK})
To check Eq.~(\ref{eq:rcacouplingK}) we first determine the last Jordan vector for the \PT-symmetric trimer in the given basis
\begin{equation}\label{eq:jbnb}
\ket{\Jordanb{\Orderb}} = \frac{1}{4\gb^2}\left(\begin{array}{c}
1/2\\
i/\sqrt{2}\\
-1/2\\
\end{array}\right)
\end{equation}
which has been chosen such that Eqs.~(\ref{eq:ortho1}) and~(\ref{eq:ortho2}) are fulfilled. As a brief consistency check the reader may verify that the Jordan vector in Eq.~(\ref{eq:jbnb}) is compatible with Eqs.~(\ref{eq:rcajordan}) and~(\ref{eq:examplercab}). Next, we determine the normalized eigenstate of the \PT-symmetric dimer
\begin{equation}
\ket{\stateEPa} = \frac{1}{\sqrt{2}}\left(\begin{array}{c}
1\\
-i\\
\end{array}\right) \ .
\end{equation}
With Eqs.~(\ref{eq:normjb}) and~(\ref{eq:exampleK}) follows 
\begin{equation}
\bra{\widetilde{\Jordanb{\Orderb}}}\oK\ket{\stateEPa} = \frac{k}{2\sqrt{2}} \ .  
\end{equation}
Using the individual response strengths in Eqs.~(\ref{eq:examplercaa}) and~(\ref{eq:examplercab}) we finally obtain from Eq.~(\ref{eq:rcacouplingK})
\begin{equation}
\rca = \sqrt{8}|k|\ga\gb^2 
\end{equation}
which is identical to Eq.~(\ref{eq:examplerca}). 

\begin{figure}[ht]
	\includegraphics[width=0.65\columnwidth]{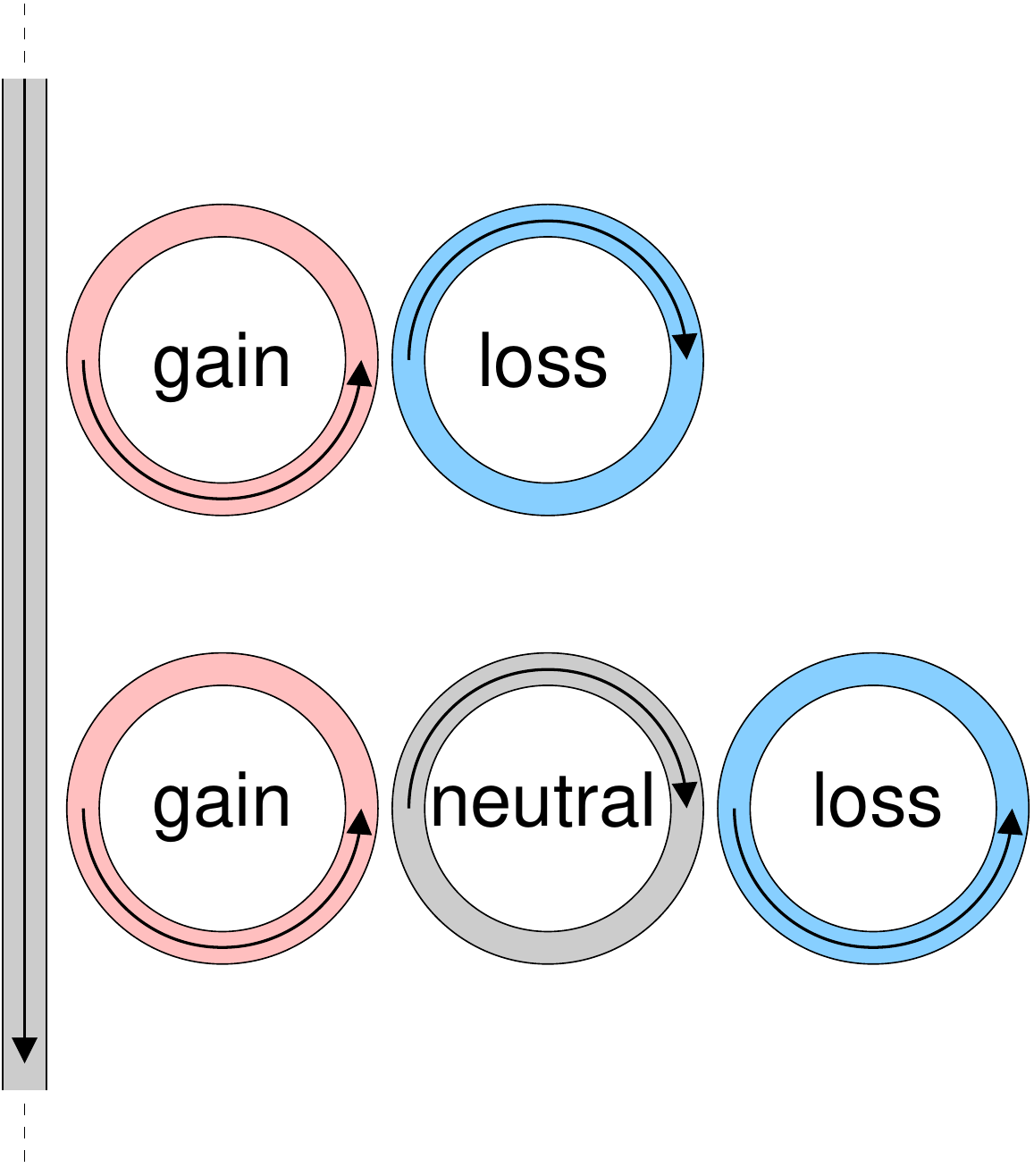}
	\caption{\HL{Illustration of a photonic implementation of the unidirectionally-coupled \PT-symmetric dimer and trimer; cf.~Fig~\ref{fig:example}. Each individual optical microring supports a CW and a CCW propagating mode. A unidirectional coupling of the modes indicated by arrows is achieved by a conventional waveguide.}}
	\label{fig:realization}
\end{figure}

\HL{
\subsection{Photonic implementation}
Figure~\ref{fig:realization} illustrates a possible implementation of the system in a realistic photonic setup. Each single-mode resonator in Fig.~\ref{fig:example} is replaced by a microring with the same gain/loss coefficient. In the relevant frequency regime each individual microring supports exactly \emph{two} modes with frequency~$\omega_0$, one propagating clockwise (CW) and one propagating counterclockwise (CCW). A CW propagating wave in a given microring couples evanescently with the above coupling strengths $\ga$ (top row) and $\gb$ (lower row) to a CCW propagating wave in a neighboring microring and vice versa. Both subsystems are coupled via a conventional optical waveguide, which allows wave propagation along both directions. In contrast to the situation in Ref.~\cite{ZKO20} the waveguide not terminated. Evanescent-coupling induced backscattering inside the waveguide and the microrings is assumed to be negligible.

% two EP5
The consequence is that this system exhibits \emph{two} uncoupled EP$_5$s: one as in  Fig.~\ref{fig:example} with the arrow at the unidirectional coupling as shown and one with the arrow reversed. The former (latter) one can be selected by exciting the system via the waveguide from above (below). It is to emphasized that this system fulfills Lorentz reciprocity. The unidirectional coupling is a result of the division of the Hilbert space into two uncoupled subspaces.
}

\HL{
% symmetry-reduced version
There is another point of view on the proposed photonic implementation in Fig.~\ref{fig:realization} for the special case of equal subsystems. For concreteness, assume that both subsystems are identical \PT-symmetric dimers. The full system possesses a mirror-reflection symmetry and the optical modes can be classified by the symmetry as even- and odd-parity modes. The odd-parity modes also appear in a symmetry-reduced version of the system with only one \PT-symmetric dimer and a semi-infinite waveguide with an end mirror. This is very similar to the system studied in Ref.~\cite{ZKO20}. Our proposed photonic implementation of the Hamiltonian in Eq.~(\ref{eq:HB1}) is therefore a generalization for nonidentical subsystems a and b. 

% imaginary gauge field
For an alternative photonic implementation one could use an imaginary gauge field realized by placing auxiliary rings with gain and loss between adjacent microrings~\cite{LGD15}. Here, the waveguide would be substituted by a single auxiliary ring placed between the two gainy microrings; cf Fig.~\ref{fig:realization}. This scheme would also lead to a doubling of the dimension of the Hilbert space and also would not require a breaking of Lorentz reciprocity. However, the experimental realization is possibly more challenging than for the one based on the conventional waveguide.
}

\subsection{Random perturbations}
\begin{figure}[ht]
\includegraphics[width=0.9\columnwidth]{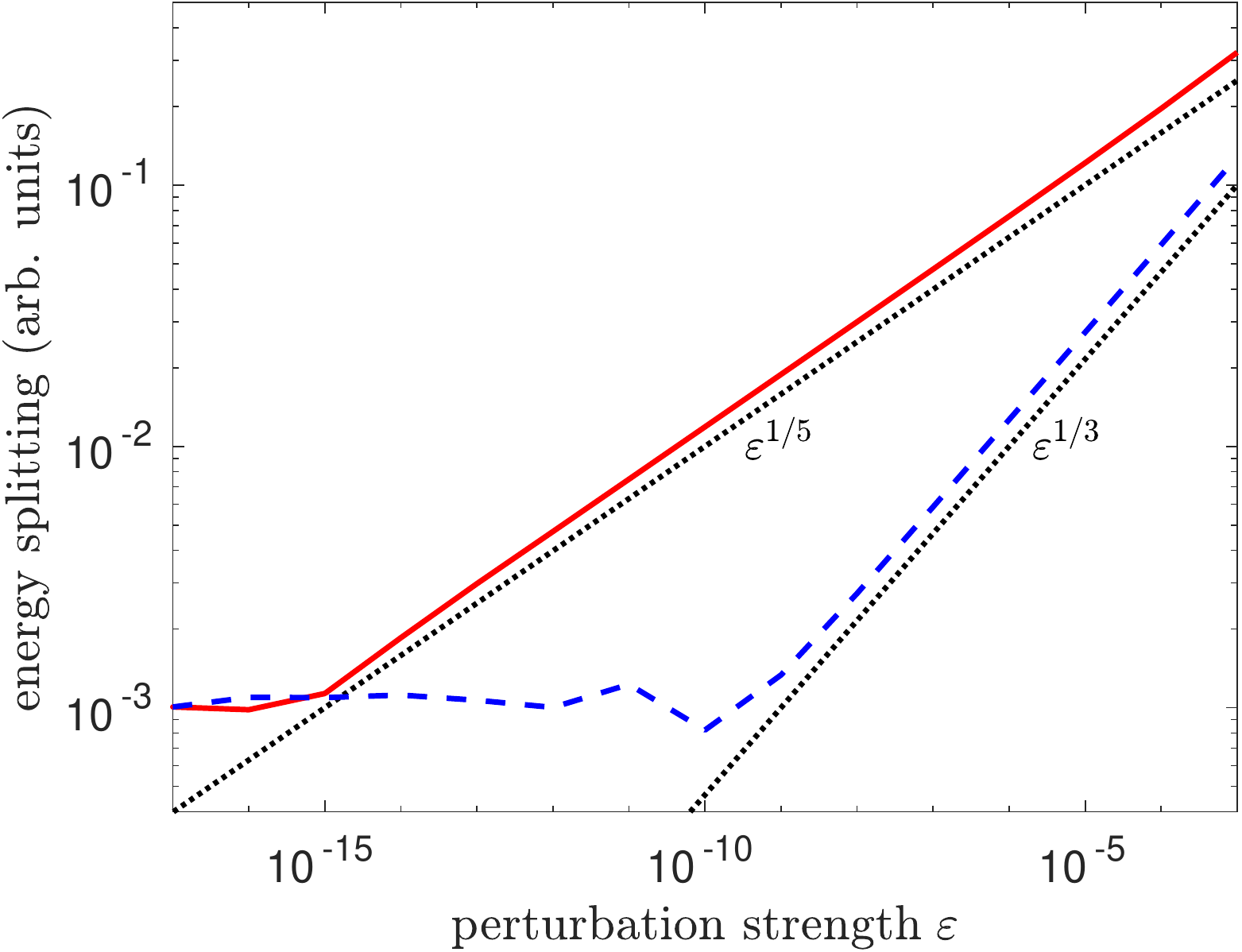}
\caption{Absolute value of the energy eigenvalue changes [$\max_j{(|E_j-\evEP|)}$ in arbitrary units] vs perturbation strength $\varepsilon$ (dimensionless) for the \PT-symmetric dimer coupled to the \PT-symmetric trimer. Note the double-logarithmic scale. The solid curve shows the case of a random generic perturbation. The dashed curve shows the case of a random unidirectional-coupling preserving perturbation. The dotted lines with slope $1/5$ and slope $1/3$ serve as a guide to the eye. The parameters are in dimensionless units $\ga = 1.5$, $\gb = 1.3$, and $k = 1$.}
\label{fig:splitting}
\end{figure}

% random generic and nongeneric perturbation
Figure~\ref{fig:splitting} shows numerical results using MATLAB for the spectral response of the full Hamiltonian in Eq.~(\ref{eq:HB1}) with Eqs.~(\ref{eq:HsPT}), (\ref{eq:H0EPd}), and~(\ref{eq:exampleK}) to a random generic perturbation and to a random unidirectional-coupling preserving perturbation. In the former case the perturbation $\Hp$ [see Eq.~(\ref{eq:H})] is chosen to be an $\order\times\order$ matrix consisting of complex random numbers with real and imaginary parts being drawn from a uniform distribution on $[-1/2,1/2]$. In the latter case the matrix elements which spoil the unidirectional coupling are set to zero, i.e., $(\Hp)_{ij} = 0$ if $i>\Ordera$ and $j \leq \Ordera$. In the generic case the splitting shows an $\varepsilon^{1/5}$ scaling as expected for an EP of order~5. In the nongeneric case, i.e., a unidirectional-coupling preserving perturbation, there is an $\varepsilon^{1/3}$ scaling corresponding to the EP$_3$ of the subsystem Hamiltonian $\Hb$.

When going to very small perturbation strengths Fig.~\ref{fig:splitting} reveals a saturation of the energy splitting around $10^{-3}$. Hence, for zero perturbation the numerical representation of the physical system is not exactly at an EP because of rounding errors due to the finite machine precision. This inaccuracy can be quantified using the inequality~(\ref{eq:specresponse}) when the rounding errors are modeled as a perturbation to the exact Hamiltonian~$\Hs$. The interpretation of $\varepsilon$ is then that of the machine precision~$\epsmp$, which for double-precision floating-point arithmetic is about $2.22\times 10^{-16}$. In this spirit $\Hp$ is considered as a random matrix where each matrix element has a zero mean and a unit variance. It is known (see, e.g., Ref.~\cite{RV10}) that the spectral norm of such an $\order\times\order$ matrix is asymptotically, i.e., for large~$\order$, given by $2\sqrt{\order}$ independent of the random distribution; for a Gaussian distribution this result is even exact. From the inequality~(\ref{eq:specresponse}) we then get the following estimate of the upper bound of the initial splitting
\begin{equation}\label{eq:splittingmp}
|\ev_j-\evEP| \leq \left(2\sqrt{\order} \epsmp \,\rca\right)^{1/n}  \ .
\end{equation}
With the parameters used in Fig.~\ref{fig:splitting}, the spectral response strength in Eq.~(\ref{eq:examplerca}), and the machine precision $\epsmp$ for double precision, we get an upper bound of the splitting of about $1.5\times 10^{-3}$ which is consistent with the data in Fig.~\ref{fig:splitting}. The observed large value of the initial splitting due to the finite machine precision is a signature of the high sensitivity of the 5th-order EP. Note that one can also use the inequality~(\ref{eq:splittingmp}) to estimate the spectral response strength by the saturated energy splitting for zero perturbation.

\section{Summary}
\label{sec:summary}
The hierarchical construction of higher-order EPs introduced in Ref.~\cite{ZKO20} has been reexamined. In this robust construction scheme two EPs of the same order are merged by a unidirectional coupling to give a single EP of twice the order. Our complementary analysis exploits the nilpotency of the traceless part of the involved Hamiltonians. It provides an alternative view on the construction scheme and allows us to generalize it to the merging of two EPs of different order (a generalization to include more than two EPs is straightforward). A condition for generic coupling matrices has been derived.

We have presented a simple formula for the spectral response strength of the resulting higher-order EP and have related it to the spectral response strengths of the original lower-order EPs. The spectral response strength is an important measure of the quality of the EP and is relevant for the design of EP-based devices.  

We have shown that unidirectional-coupling preserving perturbations of the system with EP$_\order$ are nongeneric in the sense that they do not lead to an $\varepsilon^{1/\order}$-scaling of the energy splittings with the perturbation strength~$\varepsilon$.

To illustrate our approach we have discussed the unidirectional coupling of a \PT-symmetric dimer to a \PT-symmetric trimer. For this example it has been demonstrated that the spectral response strength offers a convenient tool to estimate the rounding-error-induced energy splitting at an EP.

\HL{
The results presented here are a first step towards a broader theory of mode coupling with EPs. Here, two EPs of arbitrary order are unidirectionally coupled leading to a higher-order EP. It would be also interesting to couple two (or more) EPs in a non-unidirectional manner. This does in general not lead to an EP but it does lead to a bunch of modes with strong non-orthogonality and high sensitivity to perturbations which can be useful in particular for sensor applications.
}

\acknowledgments 
Valuable discussions with R. El-Ganainy \HL{and J. Kullig} are acknowledged. %We acknowledge support for the Book Processing Charge by the Open Access Publication Fund of Magdeburg University.

\begin{appendix}
\section{Generic energy splitting at an EP}
\label{app:splitting}
In this appendix we prove Eq.~(\ref{eq:splitting}). We consider the eigenvalue problem of the Hamiltonian in Eq.~(\ref{eq:H})
\begin{equation}\label{eq:EP}
(\Hs+\varepsilon\Hp)\ket{\state_j} = \ev_j\ket{\state_j}
\end{equation}
with eigenvalues $\ev_j$ and eigenstates $\ket{\state_j}$ normalized to unity, i.e., $\normV{\state_j} = 1$. Equation~(\ref{eq:EP}) can be written as
\begin{equation}\label{eq:EP2}
\ket{\state_j} = \varepsilon\GF(\ev_j)\Hp\ket{\state_j}
\end{equation}
with the Green's function of the unperturbed Hamiltonian 
\begin{equation}\label{eq:GFdef}
\GF(\ev) := (\ev\ident-\Hs)^{-1} \ . 
\end{equation}
Taking the inner product on both sides of Eq.~(\ref{eq:EP2}) with the eigenstate $\ket{\state_j}$ gives
\begin{equation}\label{eq:EP3}
1 = \varepsilon\sandwich{\state_j}{\GF(\ev_j)\Hp}{\state_j} \ .
\end{equation}
Up to first order in $\varepsilon$ we can write
\begin{equation}\label{eq:EP4}
1 = \varepsilon\sandwich{\stateEP}{\GF(\ev_j)\Hp}{\stateEP} \ ,
\end{equation}
with the EP eigenstate $\ket{\stateEP}$.
According to Refs.~\cite{Kato66,Heiss15,Wiersig22} the Green's function near an EP$_\order$ with eigenvalue $\evEP$ is
\begin{equation}\label{eq:GFexpansion}
\GF(\ev) = \frac{\ident}{\ev-\evEP} + \frac{\op{N}}{(\ev-\evEP)^2} + \ldots + \frac{\op{N}^{\order-1}}{(\ev-\evEP)^\order} \ ,
\end{equation}
with the nilpotent matrix $\op{N}$ from Eq.~(\ref{eq:N}).
We plug the highest-order contribution of the Green's function into Eq.~(\ref{eq:EP4}) yielding
\begin{equation}\label{eq:EP5}
(\ev_j-\evEP)^\order = \varepsilon\sandwich{\stateEP}{\op{N}^{\order-1}\Hp}{\stateEP} \ .
\end{equation}
Now, we employ an orthonormal basis $\{\ket{u_j}\}$ with $\ket{u_1} = \ket{\stateEP}$. Since $\ket{\stateEP}$ spans the image of $\op{N}^{\order-1}$, see, e.g., Ref.~\cite{Wiersig22}, we get $\bra{u_{j > 1}}\op{N}^{\order-1} = 0$. Hence,
\begin{equation}
\trace{(\op{N}^{\order-1}\Hp)} = \sandwich{\stateEP}{\op{N}^{\order-1}\Hp}{\stateEP} \ .
\end{equation}
Using this in Eq.~(\ref{eq:EP5}) finally proves Eq.~(\ref{eq:splitting}).
\end{appendix}

%\bibliography{../../../bib/fg4,../../../bib/extern}
%\bibliographystyle{aipnum4-1}

\end{document}